\documentclass[aip,graphicx]{revtex4-1}

\usepackage{url} 
\usepackage[dvipdfmx]{graphicx}
\usepackage{here}
\usepackage{comment}

\draft 

\begin{document}


\title{Alkali ion-to-neutral atom converter for the magneto-optical trap of a radioactive isotope} 



\author{Hirokazu Kawamura}
\email[]{kawamura@cyric.tohoku.ac.jp}
\affiliation{Frontier Research Institute for Interdisciplinary Sciences (FRIS), Tohoku University, Sendai, Miyagi 980-8578, Japan}
\affiliation{Cyclotron and Radioisotope Center (CYRIC), Tohoku University, Sendai, Miyagi 980-8578, Japan}

\author{Ken-ichi Harada}
\affiliation{Cyclotron and Radioisotope Center (CYRIC), Tohoku University, Sendai, Miyagi 980-8578, Japan}

\author{Tomoya Sato}
\thanks{Present address: RIKEN Nishina Center for Accelerator-Based Science, Wako, Saitama 351-0198, Japan}
\affiliation{Cyclotron and Radioisotope Center (CYRIC), Tohoku University, Sendai, Miyagi 980-8578, Japan}

\author{Saki Ezure}
\affiliation{Cyclotron and Radioisotope Center (CYRIC), Tohoku University, Sendai, Miyagi 980-8578, Japan}

\author{Hiroshi~Arikawa}
\affiliation{Cyclotron and Radioisotope Center (CYRIC), Tohoku University, Sendai, Miyagi 980-8578, Japan}

\author{Takeshi~Furukawa}
\thanks{Present address: Department of Physics, Toho University, Funabashi, Chiba 274-8510, Japan}
\affiliation{Cyclotron and Radioisotope Center (CYRIC), Tohoku University, Sendai, Miyagi 980-8578, Japan}

\author{Tomohiro Hayamizu}
\thanks{Present address: Center for Nuclear Study (CNS), University of Tokyo, Wako, Saitama 351-0190, Japan}
\affiliation{Cyclotron and Radioisotope Center (CYRIC), Tohoku University, Sendai, Miyagi 980-8578, Japan}

\author{Takeshi Inoue}
\thanks{Present address: RIKEN Nishina Center for Accelerator-Based Science, Wako, Saitama 351-0198, Japan}
\affiliation{Frontier Research Institute for Interdisciplinary Sciences (FRIS), Tohoku University, Sendai, Miyagi 980-8578, Japan}
\affiliation{Cyclotron and Radioisotope Center (CYRIC), Tohoku University, Sendai, Miyagi 980-8578, Japan}

\author{Taisuke Ishikawa}
\affiliation{Cyclotron and Radioisotope Center (CYRIC), Tohoku University, Sendai, Miyagi 980-8578, Japan}

\author{Masatoshi Itoh}
\affiliation{Cyclotron and Radioisotope Center (CYRIC), Tohoku University, Sendai, Miyagi 980-8578, Japan}

\author{Tomohiro~Kato}
\affiliation{Cyclotron and Radioisotope Center (CYRIC), Tohoku University, Sendai, Miyagi 980-8578, Japan}

\author{Akihito Oikawa}
\affiliation{Cyclotron and Radioisotope Center (CYRIC), Tohoku University, Sendai, Miyagi 980-8578, Japan}

\author{Takatoshi Aoki}
\affiliation{Graduate School of Arts and Sciences, University of Tokyo, Meguro, Tokyo 153-8902, Japan}

\author{Atsushi Hatakeyama}
\affiliation{Department of Applied Physics, Tokyo University of Agriculture and Technology, Koganei, Tokyo 184-8588, Japan}

\author{Yasuhiro Sakemi}
\thanks{Present address: Center for Nuclear Study (CNS), University of Tokyo, Wako, Saitama 351-0190, Japan}
\affiliation{Cyclotron and Radioisotope Center (CYRIC), Tohoku University, Sendai, Miyagi 980-8578, Japan}


\date{\today}

\begin{abstract}
We have developed a unique neutralizer device that uses an yttrium target surrounded by a platinum wall to magneto-optically trap radioactive atoms. 
In general, the radioactive nucleus produced in a nuclear reaction is extracted and transported in ion form. 
For the magneto-optical trap, thermal neutralization must occur on the surface of a metal with a small work function. 
The converter can produce a neutral atomic beam with small angular divergence that, given the recycling of atoms and ions, converts ions into neutral atoms with remarkable efficiency. 
We demonstrated the ion neutralization process using stable rubidium and confirmed $10^6$ neutralized atoms in the magneto-optical trap. 
Additionally, the experiment using francium demonstrated the obtaining of neutralized francium atoms. 
\end{abstract}

\pacs{}

\maketitle 

\section{Introduction}\label{sec1}
The violation of fundamental symmetries significantly affects fundamental physics. 
In atomic systems, the effects extend to the electric dipole moment and parity non-conservation effect, which can be enhanced by using a heavier element \cite{GINGES200463, 0953-4075-43-7-074004, Aoki2017}. 
Exceptionally heavy elements that lack stable isotopes have to be produced artificially, so that what is observed during experiments can be measured. 
In contrast, laser cooling and trapping techniques can better aid such precise experiments. 
Among these, a magneto-optical trap (MOT) is a standard technique to cool and trap neutral atoms. 

Trapping radioactive atoms produced via accelerators has been studied for various species of atoms by many researchers since ${}^{21}\mathrm{Na}$ \cite{PhysRevLett.72.3791} and ${}^{79}\mathrm{Rb}$ \cite{PhysRevLett.72.3795} were trapped in 1994. 
Locating a trapping area far from the production area, where radioactivity is exceptionally high, requires the long-distance transport of ions. 
Researchers have generally recommended thermally neutralizing radioactive ions on the surface of a metal with a small work function, such as zirconium (Zr) \cite{PhysRevC.79.015501, DiRosa2003, PhysRevA.97.042507} or yttrium (Y) \cite{PhysRevLett.72.3795, PhysRevLett.76.3522}. 
That process can convert an ion beam in the keV energy range into neutral atoms with less than 1 eV of energy. 
The neutralizer should be surrounded by a coated-glass cell so that desorbed neutralized atoms are not lost \cite{PhysRevA.58.R1637,    DiRosa2003,       doi:10.1063/1.1606093,     PhysRevA.78.063415}. 
Precooling --- by, for example, transverse cooling and/or Zeeman slowing \cite{doi:10.1063/1.2069651,        1367-2630-12-6-065031} --- can reduce the atomic velocity from a few hundred m/s to a few tens of m/s, thus greatly improving trapping efficiency. 
However, neutralizers are unavailable for precooling techniques. 

In some cases, the neutralization of radioactive ions is necessary, for which a charge exchange with alkali vapor can create an atomic beam of radioactive alkali isotopes \cite{LEVY2002253,        doi:10.1063/1.3271037}. 
Although this method affords high conversion efficiency for fast (several keV) radioactive ions, it is difficult to apply precooling and trapping techniques for a fast atomic beam when neutralized. 
Neutralization using yttrium can aid the optical spectroscopy of neutral atoms \cite{PhysRevLett.76.3522,       TOUCHARD1981329,       Stroke2006} as well as the measurements of nuclear magnetic moments, via the atomic beam magnetic resonance method \cite{PINARD20053,     NAGAE2005580}. 
These neutralization methods can produce partially-directional radioactive atoms. 

In a novel approach to trapping a short-lived radioactive alkali element offline without an accelerator, Lu et al. achieved a highly-efficient MOT of ${}^{221}\mathrm{Fr}$ by using an orthotropic source, as described below \cite{doi:10.1063/1.1146804,         PhysRevLett.79.994}. 
However, their setup cannot accommodate continuous, long-term experiments, because the accelerator cannot supply radioisotopes. 

In this paper, we report the development of a unique ion-to-atom converter that can produce an atomic beam with small angular divergence and thermal distribution, and can be used for transverse cooling and/or Zeeman slowing. 
Using an orthotropic source, the converter can be used for long-term experiments with accelerators. Ongoing experiments have been using an AVF cyclotron to trap radioactive Fr atoms after Fr ions were neutralized \cite{KAWAMURA2013582,      Inoue2015,     Harada:16}. 
We conducted experiments that used reionization detection and an MOT with a stable rubidium (Rb) isotope, in order to confirm the neutralization of ions. 
Neutralized Rb atoms were observed, and $10^6$ atoms were trapped in the MOT. 

\section{Experiment}
\begin{figure}[htbp]
  \centering
  \includegraphics[width=7cm]{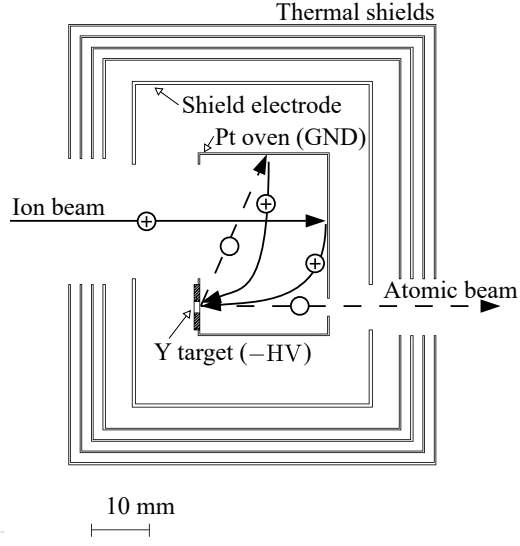}
  \caption{Ion-to-atom converter based on the orthotropic source. 
  The shaded parts next to the Y target are ceramic insulators. 
  HV and GND indicate high voltage and ground, respectively. 
  }
  \label{fig:Francois}
\end{figure}

\begin{figure*}[htbp]
  \begin{minipage}{0.47\hsize}
  \includegraphics[width=\textwidth]{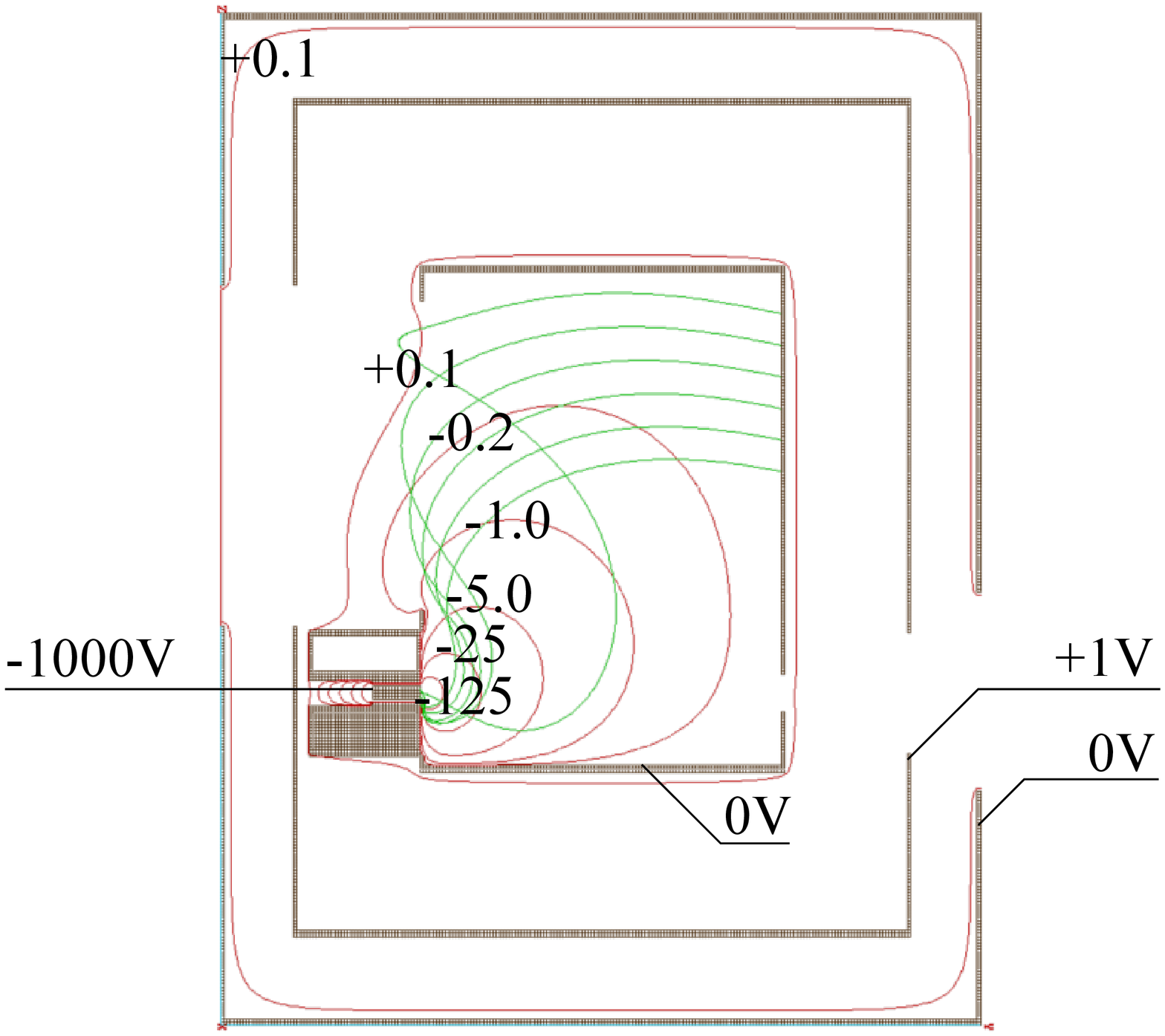}
  \end{minipage}
  \begin{minipage}{0.49\hsize}
  \includegraphics[width=\textwidth]{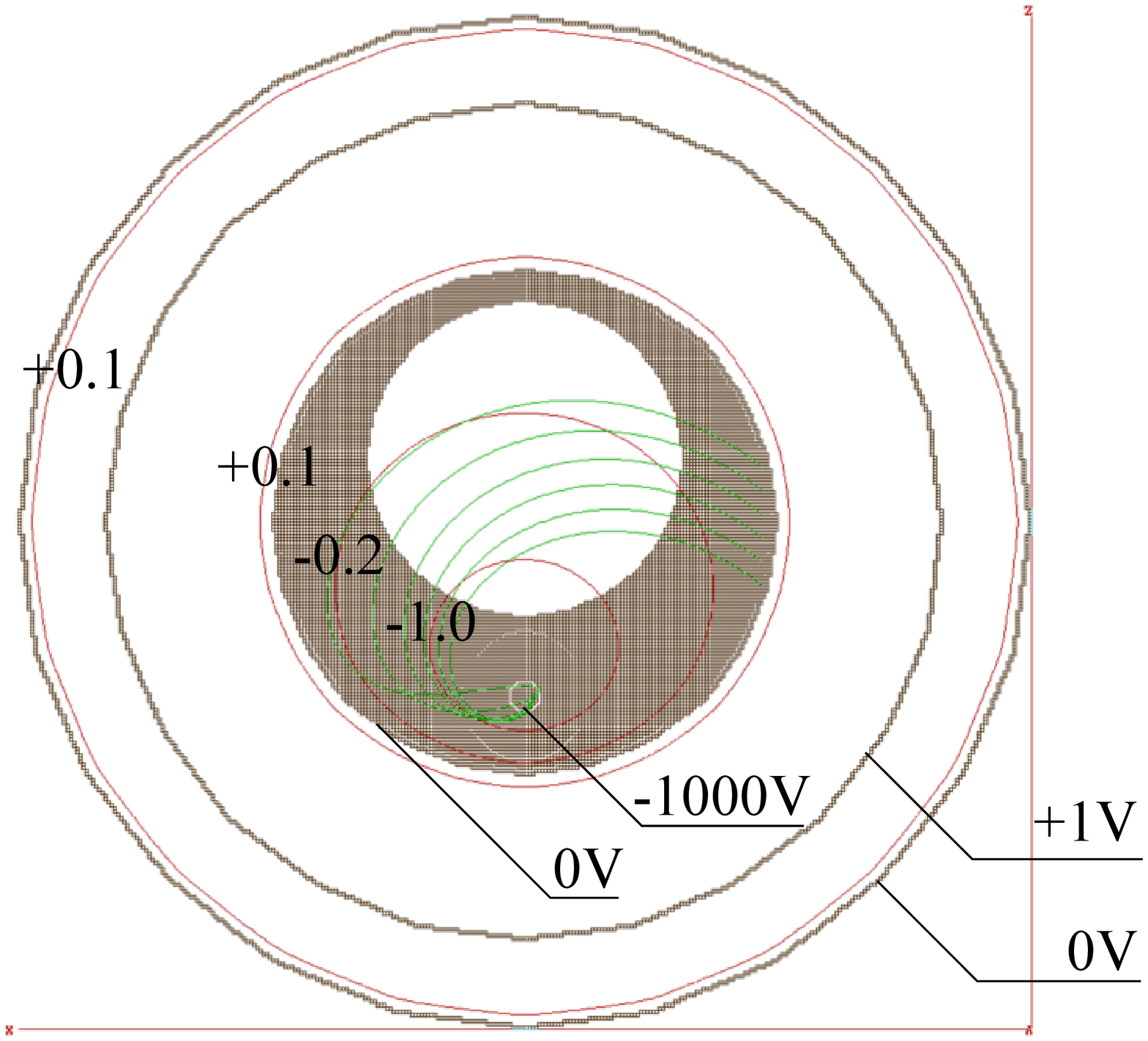}
  \end{minipage}
  \caption{Results of the electric field calculation with SIMION. 
  The left and right figures are sectional views, parallel and perpendicular to the beam direction, respectively. 
  The positive ions (shown by green curves) produced at 0.2 eV from the inner wall of the oven were attracted by the Y target. 
  The voltages of the Y target, Pt oven, shield electrode, and thermal shields are $-1000$ V, $0$ V, $+1$ V, and $0$ V, respectively. 
  Red curves show equipotential planes, and the numbers on the curves denote their potentials in units of V.
}
  \label{fig:simion}
\end{figure*}

An orthotropic source is an experimental device that exploits thermal ionization and neutralization \cite{doi:10.1063/1.1146804}. 
Our setup included a neutralizer target and an ionizer oven (Kitano Seiki, Tokyo) \cite{KitWEB} in the same manner as the original source \cite{doi:10.1063/1.1146804} (Figure \ref{fig:Francois}). 
The target and the oven were made using tantalum coated with Y (100 nm thickness and 99.9\% purity) and Pt (100 nm thickness and 99.99\% purity), respectively. 
The oven has an aperture for the output of the atomic beam, as well as a large hole for the input of the ion beam. 
The dimensions are as follows: a 2-mm diameter target, a 3-mm diameter aperture at a distance of 23 mm from the target, and a 20-mm diameter hole. 
The diameter of the hole for the beam input was determined based on the spot size of our ion beam. 
The other dimensions were determined by investigating conditions that would maximize the output beam intensity with Monte Carlo simulation, based on the dimensions of the original source \cite{doi:10.1063/1.1146804}. 
Around the oven, an additional electrode produced an electric field to confine the ions. 
However, the electrode's confinement function was secondary to its function as a thermal shield. 
These electrodes were evaluated with SIMION code \cite{SIMION8}, to ensure that the ions were attracted to the Y target, by the electric field they generated (Figure \ref{fig:simion}). 
Additionally, as a thermal shield, the electrode was surrounded by four additional thermal shields that were electrically grounded. 
When incoming ions entered the wall of the Pt oven, the beam current could be read from the current that flowed from surface charges on the oven at room temperature (as it floated electrically, similar to a Faraday cup). 

As ions entered the heated Pt oven, they were emitted from the oven and thereafter attracted to the surface of the Y target, to which a negative electric voltage was applied. 
According to Dinneen et al. \cite{doi:10.1063/1.1146804}, the abundance of alkali atoms and ions on a high-temperature metal surface can be represented by the Saha-Langmuir equation \cite{Langmuir61,      PhysRev.48.960}:  
\begin{equation}\label{eq:sahalangmuir}
\frac{n_+}{n_0} = \frac{1}{2}\exp\left( \frac{\phi-E_i}{k_BT} \right) 
\end{equation}
in which $n_+/n_0$ represents the ratio of ions to atoms with ionization energy $E_i$ evaporated from a surface of work function $\phi$ at a temperature $T$, and $k_B$ is the Boltzmann constant. 
This equation is often used to calculate surface \textit{ionization} even for heavy elements \cite{Sato2015}, in which the smaller work function of the metal target can prompt a more efficient \textit{neutralization}. 
The $E_i$ of the alkali elements are 4.1 eV for Fr and 4.2 eV for Rb \cite{Lid03}, whereas the $\phi$ for Y is 3.1 eV \cite{Lid03}. 
Consequently, the ions stuck to the Y target converted to neutral atoms by accepting electrons from the target surface, due to the process represented in Eq. (\ref{eq:sahalangmuir}). 
When the Y target surface was heated, the neutral atoms were desorbed from the target, which characterizes the process as a source of neutral atoms; the atoms then passed through the aperture of the oven and exited via the finite output angle. 
Otherwise, atoms interrupted by the wall of the oven became ionized on the surface of Pt, given its large $\phi$. 
When the ionized particles were attracted by the Y target, the particles neutralized on the Y surface desorbed into the oven. 
Due to those cycles of ionization and neutralization, an atomic beam with small angular divergence is produced, minimizing the loss of the radioactive atoms. 

\subsection{Reionization detection}
\begin{figure*}[t]
  \centering
  \includegraphics[width=\textwidth]{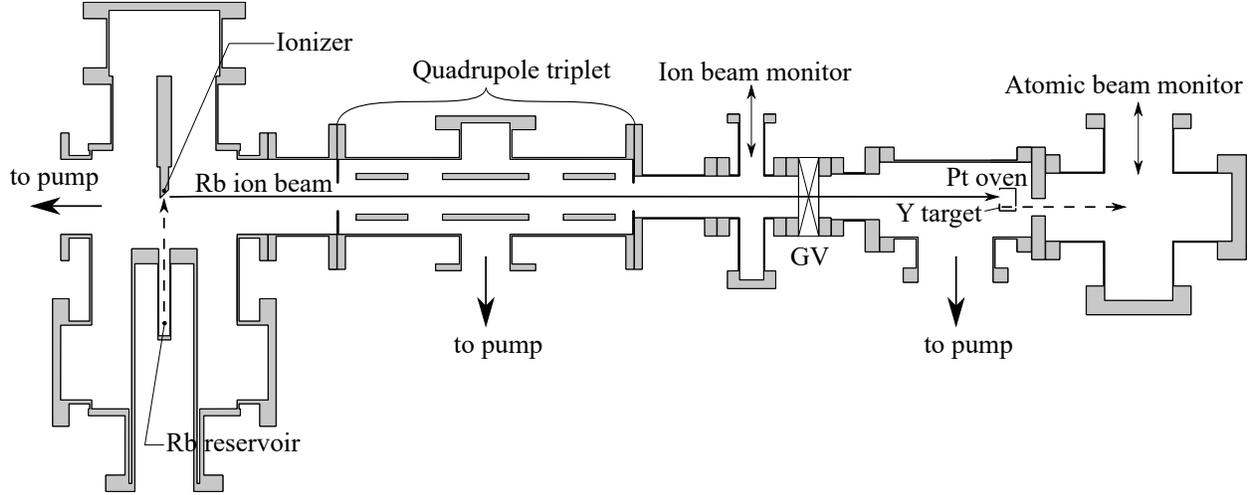}
  \caption{Overall view of the apparatus, from the Rb ion source to the converter and the atomic beam monitor. 
  The Rb ions were produced through thermal ionization of Rb vapor on a heated Mo target \cite{Sato11}. 
  The ion beam was focused within an electrostatic quadrupole triplet lens and was transported. 
  The ion beam monitor, placed after the triplet lens, was able to measure the ion beam current. 
  A gate valve (GV) can stop the beam and decouple vacuum conditions. 
  After the GV were the converter and the atomic beam monitor, which is the reionization detector. 
}
  \label{fig:RbIS_to_Francois}
\end{figure*}

\begin{figure}[htbp]
  \centering
  \includegraphics[width=8.5cm]{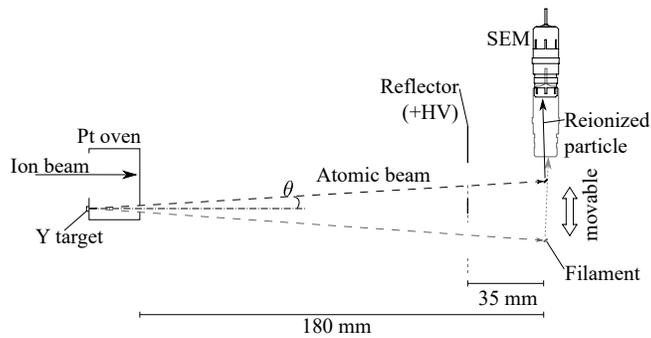}
  \caption{Setup of the reionization detector, which can measure the angular $\theta$ dependence of the beam intensity. 
  SEM indicates the secondary electron multiplier (5901 magnum electron multiplier, Photonis, Sturbridge, MA, US). 
  }
  \label{fig:CEM}
\end{figure}

Experiments using reionization detection were performed in order to confirm that neutralized atoms can be produced with the ion-to-atom converter. 
Figure \ref{fig:RbIS_to_Francois} shows the overall view of the apparatus, from the Rb ion source to the converter and the reionization detector. 
The energy of the Rb beam was typically 1.5 keV. 
Figure \ref{fig:CEM} presents a schematic view of our reionization detector, in which the rhenium filament (dimensions 10 mm $\times$ 0.8 mm $\times$ 25 $\mu$m) can convert neutral atoms into ions detectable by a secondary electron multiplier. 
An ion reflector (to which a suitable voltage was applied) was placed in front of the filament, which could only be reached by neutral components, since ions were swept out by the reflector through an opening 16 mm in diameter. 
Altogether, the reionization detector allowed the identification of neutral components emitted from the active ion-to-atom converter. 
The counting rate of the reionization detector increased as the oven temperature increased, as monitored by a WRe5/26 thermocouple located on the underside of the Pt oven. 
The measured temperature was kept within $\pm1^{\circ}$C during each run, and signals were observed at temperatures greater than $700^{\circ}$C. 

\begin{figure}[htbp]
  \centering
  \includegraphics[width=7cm]{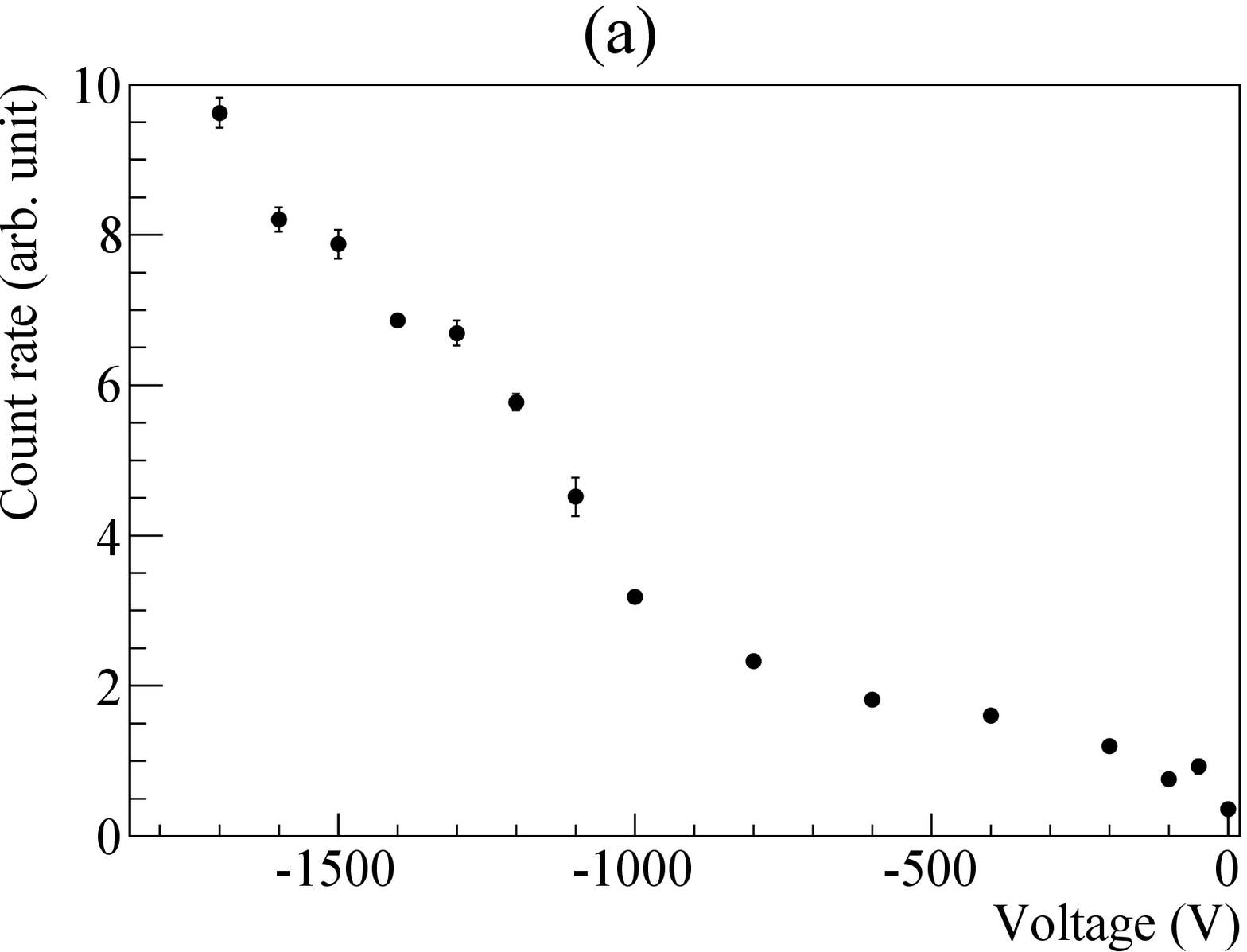}
  \includegraphics[width=7cm]{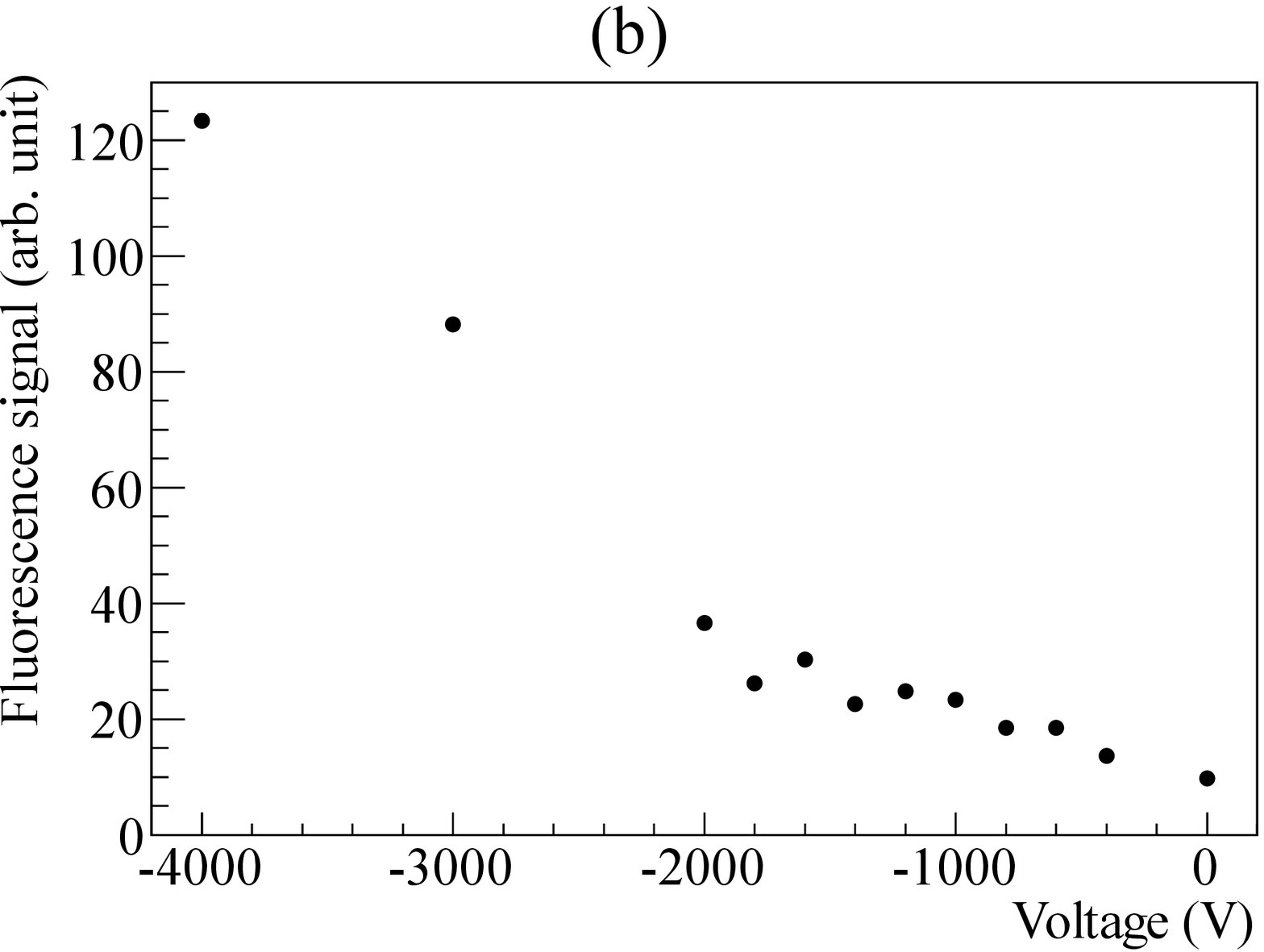}
  \caption{(a) Attracting-neutralizer voltage dependence for the counting rate of reionization detector. 
  (b) The fluorescence signal of trapped atoms as a function of the attracting voltage. 
  The largest uncertainty would be a systematic error originating from the fluctuation of the laser frequencies, but it was not manifested. 
}
  \label{fig:VolDep}
\end{figure}

The dependence of the counting rate on the Y-target voltage was measured at an oven temperature of $1000^{\circ}$C (Figure \ref{fig:VolDep} (a)). 
The counting rate increased as the negative voltage increased --- although some of the ions may have escaped from the oven, as the attracting field generated by the Y target was not completely able to attract atoms in the distance, which suggests that strengthening the field of attraction can reduce the number of escaping ions and can consequently increase atom output. 
However, the output efficiency has to be saturated; otherwise, all ions are attracted at an exceptionally high voltage.

\begin{figure}[htbp]
  \centering
  \includegraphics[width=7.5cm]{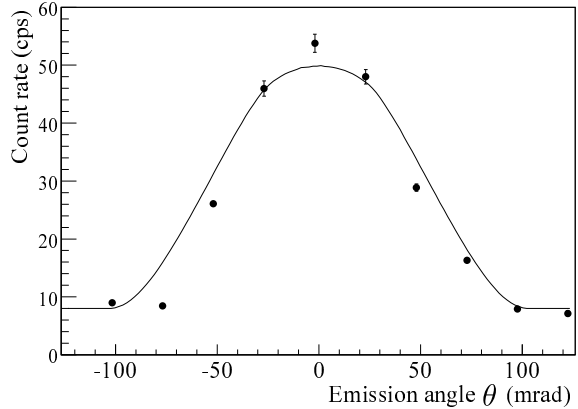}
  \caption{
  Emission angle $\theta$ distribution of the output atomic beam. The curve shows the expected angular distribution based on the geometry of the converter and detector. 
}
  \label{fig:angular}
\end{figure}

Figure \ref{fig:angular} shows the broadening of output atoms at $970^{\circ}$C. 
The output should be geometrically limited by the dimensions of the converter, which would produce an atomic beam with smaller angular divergence. 
The reionization detector can be moved transversely, in relation to the beam direction (Figure \ref{fig:CEM}). 
The angular $\theta$ dependence of the beam's intensity was derived from the position dependence, and the atoms' output was ultimately based on the angular limitation. 
The width of such a broadening corresponded to approximately 100 mrad, which was less than that measured by Dinneen et al. \cite{doi:10.1063/1.1146804}. 
The curve in Figure \ref{fig:angular} shows the number of events (which is limited by the dimensions of the converter's aperture and the reionization detector's filament), where all events are random straight lines through the Y target. 
This implies that the beam broadening can almost be explained by the geometry of the equipment. 

\begin{figure}[htbp]
  \centering
  \includegraphics[width=8cm]{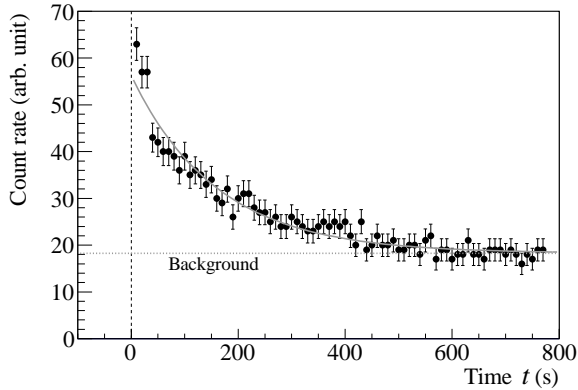}
  \caption{Time evolution for counting rate after the ion beam stops. These error bars are dominated by reading errors. 
}
  \label{fig:ResTime}
\end{figure}

Figure \ref{fig:ResTime} displays the amount of time the atoms remained in the oven at a temperature of $700^{\circ}$C. 
The ion beam was continuously supplied in the oven until the gate valve closed at $t=0$ s. 
The counting rate of the reionization detector decayed exponentially over time. 
The process (Figure \ref{fig:ResTime}) provides the mean time during which ions and neutral atoms remaining in the oven are derived amid cycles of ionization and neutralization. 
The time constant derived from an exponential fitting was $160\pm15$ s. 
Furthermore, we discuss the cycle time and the recycle probability, based on this decay curve, using a simplified model of particle behaviors. 
Nearly all particles entering the device were assumed to remain there, due to the recycling atoms and ions. 
In our model, a \textit{cycle} refers to the sequence in which a particle desorbs from an inner surface of the device and stops at another surface. 
With $\varepsilon$ as the probability that the particle strays from the cycle, and $t_1$ as the duration of one cycle, the time evolution $N(t)$ of the number of particles remaining inside the device can be expressed as:
\begin{equation}\label{eq:timeconstant}
N(t) = C_0 ( 1 - \varepsilon)^{t/t_1} + C_1 
\end{equation}
in which $C_0$ and $C_1$ indicate constants. 

Applying Eq. (\ref{eq:timeconstant}) to the process shown in Figure \ref{fig:ResTime} yields 
$\varepsilon = (1.2\pm0.3)\times10^{-3}$ and $t_1 = (0.17\pm0.04)$ s. 
$t_1$ includes both the time of flight of the particles (which is quite brief, due to the particles' high velocity) as well as the much longer duration of surface diffusion and/or desorption. 
In this experiment, we should consider the situation that Rb ions accelerating by a few kV enter the surface of Pt or Y. 
The diffusion constant $D_{700^{\circ}\mathrm{C}}$ is predicted as $10^{-12}$ -- $10^{-14}$ $\textrm{cm}^{2}\textrm{s}^{-1}$ by the references \cite{MELCONIAN200593, PhysRevA.78.063415}. 
Since the implantation depth is a few nm, the time to reach the surface is less than a few seconds. 
This time is consistent with the derived cycle duration $t_1$, as well as with the evaporation time ($<1$ s at $700^{\circ}$C) mentioned in a reference \cite{doi:10.1063/1.1146804}. 
Since the desorption time decreases as the surface temperature increases, higher temperatures will shorten the output time. 
Moreover, since $\varepsilon$ encompasses particles escaped through the ion entrance, $\varepsilon$ will be partly reduced if the hole size of the entrance is reduced. 

\subsection{Detection in a magneto-optical trap}
Reionization detection cannot guarantee that the detected component is a reliably neutral Rb atom, because the method does not allow users to identify particles. 
In fact, other alkali elements, such as impurities among the device materials, could behave similarly to the Rb beam \cite{PhysRev.35.381}. 
However, magneto-optical trapping is an atom-trapping technique that uses the transition of a hyperfine structure with a laser light, and as such, it can allow precise identification of the Rb isotope. 

\begin{figure*}[htbp]
  \centering
  \includegraphics[width=13cm]{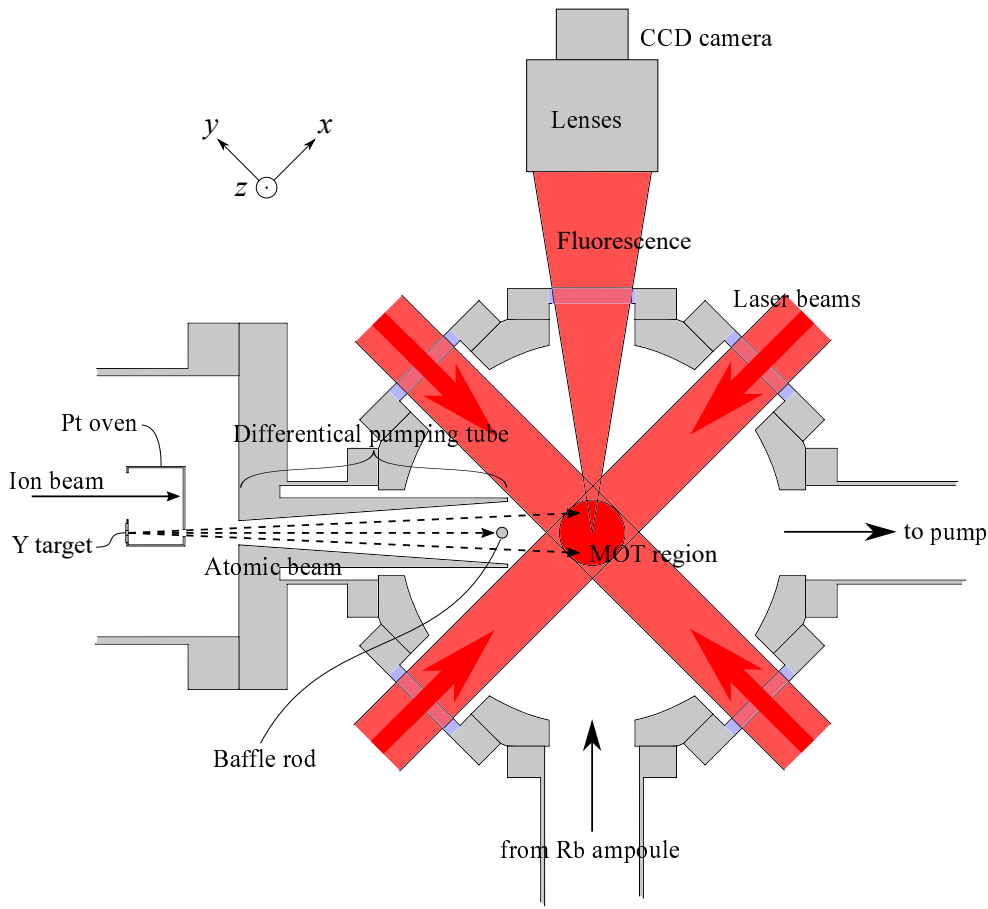}
  \caption{Setup of the magneto-optical trap. The direction orthogonal to the page is the $z$ axis. 
  Not only the converter but also the Rb ampoule holder can be found just around the MOT chamber. 
  This ampoule was able to directly supply neutral Rb atoms for pilot experiments. 
}
  \label{fig:MOTsetup}
\end{figure*}

We performed the magneto-optical trapping of Rb atoms \cite{KAWAMURA2013582} after converting the reionization detector into the MOT chamber, as shown in Figure \ref{fig:MOTsetup}. 
Although the vacuum pressure of the converter was $4\times10^{-5}$ Pa, due to its high temperature, 
the background pressure of the MOT chamber remained at $8\times10^{-7}$ Pa, due to the differential pumping duct tube, which has a 10-mm diameter on the side of the converter, a 26-mm diameter on the side of the MOT, and a length of 109.85 mm. 
A 3-mm diameter baffle rod attached to the end of the tube stopped a portion of the outgoing atoms, in order to prevent trapped atoms from colliding with fast ones. 
The octagonal chamber was made of stainless steel, and the viewing ports were coated with antireflective material. 
The trapping light was formed by three retroreflector laser beams with diameters of 30 mm each. 
The typical power of the trapping beam was 18 mW per arm, and the frequency of the beam was detuned by $-20$ MHz from the center of the resonance ${}^{87}$Rb $5S_{1/2}$ $(F=2)$ $\to$ $5P_{3/2}$ $(F'=3)$. 
$F$ ($F'$) to represent the total angular momentum of the ground (excited) state. 
The repumper light passed through the same optics as the $x$ and $y$ axes of the trapping light. 
The total power of the repumper beams on two axes was 8 mW, and the frequency was tuned to approximately the value of the resonance of the transition $5S_{1/2}$ $(F=1)$ $\to$ $5P_{3/2}$ $(F'=2)$. 
A current of 15 A, running through a pair of 166-mm diameter coils in anti-Helmholtz configuration, generated a magnetic field gradient of 30 Gauss/cm along the $z$ axis.

The experiment demonstrated a method to trap neutral Rb atoms originating from the Rb ion beam. 
Beforehand, trapped Rb atoms were not observed without an ion beam, after it was confirmed that the trapping system was working normally with the Rb ampoule for the pilot test. 
In other words, the trapped Rb atoms that were observed during the supply of the ion beam originated from the ion beam. 
The temperature of the converter can be adjusted to maximize the number of trapped atoms. 
When the converter temperature rises, the number of the output atoms should increase while the trapping efficiency should decrease, because the atom velocity and the background pressure must increase. 
Although an optimum temperature could not be found at significant levels, an empirically-supported temperature was used in this work. 
Figure \ref{fig:VolDep} (b) presents the Y-target voltage dependence of the fluorescence signal, taken with a CCD camera, at an oven temperature of $1010^{\circ}$C. 
The fluorescence signal should be proportional to the number of trapped atoms and should increase with raised voltage, as in the reionization detection.

In the experiments, we trapped a maximum of $n=10^6$ atoms when the ion beam current (including Rb and background elements) was 10 $e$nA. 
This corresponds to the situation from which the maximum fluorescence signal was derived in Figure \ref{fig:VolDep} (b). 
When the number of trapped atoms was estimated, it assumed that the number of background atoms was zero because background components in trapped atoms were not measured. 
The intensity $I$ of ${}^{87}$Rb ions was estimated to be roughly $10^{10}$ ions/s from its abundance ratio. 
The relationship between the input ion intensity $I$ and the number of trapped atoms $n$ can be expressed as
\begin{equation}
n = I  \eta  w  \tau 
\end{equation}
in which: $\eta$ indicates the conversion efficiency of the total output of neutralized atoms to the input ions; 
$w$ indicates the proportion of atoms at a velocity slower than the capture velocity to the total atoms; 
and $\tau$ indicates the lifetime of trapped atoms. 

Assuming that the photon scattering rate is saturated, 
we defined a capture velocity $v_c$ as the velocity that becomes 0 by deceleration due to the maximum scattering force $ma=\hbar k \gamma/2$ after the motion of the distance $r$. 
Accordingly, $v_c$ can be estimated as $v_c = \sqrt{r\hbar k \gamma/m}$ from a simplified equation of motion, 
in which $r$ is the radius of trapping light, $k$ is the photon wave vector, $\gamma$ is the photon scattering rate, and $m$ is the mass of ${}^{87}$Rb. 
When $r=15$ mm, $v_c$ is estimated to be 60 m/s. 
If the neutralized Rb has a Maxwell-Boltzmann velocity distribution at an oven temperature of $1010^{\circ}$C, 
then the most probable velocity is 500 m/s, and the trappable fraction $w$ would be 0.1\%. 
Consequently, the efficiency $\eta$ is 10\% under the assumption that $\tau$ is 1 s. 
If the recycling of atoms and ions functions perfectly, then $\eta$ must be 100\%. 
This result implies that some particles escape from the device, whereas others never desorb from the inner wall. 
Additionally, $\eta$ in our experiment included the effect that some components of the output atoms stopped at the baffle rod. 

Finally, we performed an experiment using radioactive Fr (${}^{209}$Fr, ${}^{210}$Fr, and ${}^{211}$Fr) and obtained $\eta=0.06\%$ with an $\alpha$-ray spectroscopy \cite{refId0}. 
The incident beam conditions of Rb and Fr were almost the same, except for the ion beam intensity. 
The beam intensity of the incident Fr ion was deduced to be approximately 2000 ions/s. 
This efficiency was quite small, even after considering that the Fr are short-lived isotopes (their half-lives are $t_{1/2}^{\mathrm{Fr}\mathchar`- 209}=50$ s, $t_{1/2}^{\mathrm{Fr}\mathchar`- 210}=191$ s, and $t_{1/2}^{\mathrm{Fr}\mathchar`- 211}=186$ s). 
We propose that the difference between the results stems from the deterioration of the surface conditions of the Pt and Y, although more detailed investigations of surface conditions are necessary. 
In the studies on the neutralization of different materials, the results suggested that the neutralization efficiency could be changed by a beam irradiation or by oxidation of the neutralizer \cite{Aoki16}. 
It seems that the efficiency change reflected the changes in the surface condition and the effective work function. When $\eta=0.06\%$ was obtained, the converter had already experienced the Rb experiments; 
in addition, the neutralization efficiency of Rb ions had seemed to decrease. 
It is important for future work to employ a brand-new and clean target to reach a high neutralization efficiency. 

\section{Conclusion}
We developed an orthotropic-source type converter and demonstrated the neutralization of Rb ions and the magneto-optical trapping of neutral Rb atoms. 
The conversion efficiency $\eta$ was estimated to be roughly 10\% of the oven temperature $1010^{\circ}$C. 
The converter can improve the trapping efficiency by way of transverse cooling and/or Zeeman slowing techniques, given its ability to produce an atomic beam with small angular divergence, unlike conventional neutralizers. 
Some particles, including high-velocity ions and atoms inadequately neutralized at Pt, escaped from the ion entrance of the oven, which marks the greatest difference of this study from the original source \cite{doi:10.1063/1.1146804}. 
Conversion efficiency could also be improved by improving the geometrical structure, in terms of the size of the ion entrance of the oven and the distance from the inner wall of the oven to the neutralizer target. 
The goal of creating the converter was to use it in the application of radioisotopes. 
The experiment neutralized the Fr ion (produced with an accelerator) and the neutralized Fr atom was obtained, though the conversion efficiency $\eta$ was 0.06\%. 
Unstable isotopes with long lifetimes can likely be converted, given their long durations in the oven. 
In the future, we plan to devise a more efficient laser cooling and trapping method for unstable elements, based on the development of the converter itself and the cooling of the atomic beam. 

\section*{Acknowledgment}
This work was supported by Grants-in-Aid for Scientific Research (Nos. 21104005, 23740166 and 25610112) and Tohoku University's Focused Research Project Fund.

\end{document}